\documentclass[12pt]{iopart}

\usepackage{graphics}
\usepackage[hidelinks]{hyperref}
\hypersetup{
    colorlinks=true,
    linkcolor=blue,
    filecolor=blue,      
    urlcolor=blue,
    citecolor=blue,
}
\usepackage[dvipsnames]{xcolor}
\usepackage{xcolor}
\definecolor{burgundy}{rgb}{0.5, 0.0, 0.13}
\definecolor{britishracinggreen}{rgb}{0.0, 0.26, 0.95}
\definecolor{denim}{rgb}{0.08, 0.38, 0.74}
\definecolor{deepcarmine}{rgb}{0.66, 0.13, 0.24}
\newcommand{\juan}[1]{{\color{denim}{{#1}}}}

\begin{document}

\title[Evolving disorder and chaos enhances the wave speed of elastic waves]{Evolving disorder and chaos enhances the wave speed of elastic waves}

\author{M. Ahumada$^{1, 2}$, L. Trujillo$^3$ and J. F. Mar\'in$^4$}

\address{$^1$ Departamento de F\'isica, Universidad T\'ecnica Federico Santa Mar\'ia, Casilla 110 V, Valpara\'iso, Chile.}

\address{$^2$ Departamento de F\'isica, Facultad de Ciencia, Universidad de Santiago de Chile, Usach, Av. V\'ictor Jara 3493, Estaci\'on Central, Santiago, Chile.}

\address{$^3$ AIstroSight, Centre Inria de Lyon, Hospices Civils de Lyon, Université Claude Bernard Lyon 1, F-69603 Villeurbanne, France}

\address{$^4$ Departamento de Física, Facultad de Ciencias Naturales, Matemática y del Medio Ambiente, Universidad Tecnológica Metropolitana, Las Palmeras 3360, Ñuñoa 780-0003, Santiago, Chile.}

\ead{\href{mailto:j.marinm@utem.cl}{j.marinm@utem.cl}}

\begin{abstract}
Static or frozen disorder, characterised by spatial heterogeneities, influences diverse complex systems, encompassing many-body systems, equilibrium and nonequilibrium states of matter, intricate network topologies, biological systems, and wave-matter interactions. While static disorder has been thoroughly examined, delving into evolving disorder brings increased intricacy to the issue. An example of this complexity is the observation of stochastic acceleration of electromagnetic waves in evolving media, where noisy fluctuations in the propagation medium transfer effective momentum to the wave. Here, we investigate elastic wave propagation in a one-dimensional heterogeneous medium with diagonal disorder. We examine two types of complex elastic materials: one with static disorder, where mass density randomly varies in space, and the other with evolving disorder, featuring random variations in both space and time. Our results indicate that evolving disorder enhances the propagation speed of Gaussian pulses compared to static disorder. Additionally, we demonstrate that the enhanced speed effect also occurs when the medium evolves chaotically rather than randomly over time. The latter establishes that evolving randomness is not a unique prerequisite for observing the enhanced transport of wavefronts, introducing the concept of chaotic speed enhancement of waves in complex media.
\end{abstract}

%
\vspace{2pc}

\vspace{1cm}
%
\noindent{\submitto{\JSTAT}}
%
%
%
\newpage

\tableofcontents 

\section{Introduction}

A heterogeneous medium, characterised by spatial variations in physical properties, stands in contrast to homogeneous counterparts. These variations originate from localised
changes in the medium's properties and can be effectively modelled using continuous, piecewise, or random spatial functions~\cite{Sahimi2003,bahraminasab2007}. Found extensively in nature, materials exhibiting such inherent heterogeneity, including granular materials, porous media, glassy systems, the atmosphere, and colloidal suspensions, play crucial roles in technological and industrial applications influenced by human manipulation~\cite{deGennes1999}. Efforts to model complex materials have led to stochastic descriptions of processes within these systems~\cite{bahraminasab2007, Buryachenko2007, Christakos2012}. These spatial random field models find applications in various real-world phenomena, such as transport processes in porous media~\cite{Quintard1993}, oil reservoir characterisation parameters~\cite{Sahimi1992, Wu2006}, and the spatial distribution of elastic moduli in polycrystals~\cite{Hamanaka2006}, granular materials \cite{Aoki1996, Luding2001, Trujillo2010}, and composites~\cite{Mohanty1982, Torquato1985, Shalaev1999}. Numerical modelling has gained significant interest for practical applications, with acoustic and elastic wave propagation analysis serving as a key method for non-invasive tomography and seismic wave source characterisation in geophysics~\cite{Boore1972, Kelly1976, Frankel1984, Frankel1986, Dablain1986, Kneib1993}.

Wave localisation, a robust phenomenon observed in various disordered systems with spatially dependent randomness, i.e. frozen disorder \cite{Lagendijk2009}, has been a subject of significant exploration~\cite{he1986,maynard2001,sahimi2009}. Phillip Anderson's seminal work in 1958 demonstrated that frozen disorder can transform a conductor into an insulator, halting transport due to multiple scattering~\cite{Anderson1958}. This disorder-induced phase transition, characterised by a change from a classical diffusion regime to a localized state, appears in the transport phenomena of both strongly and weakly interacting particles, including light in disordered photonic crystals \cite{Wiersma1997, Iomin2015}, water waves \cite{Devillard1988, Belzons1988, Ricard2024} and acoustics \cite{Condat1987, Figotin1996}. On the other hand, environmental noise and spatial disorder are ubiquitous in nature, and although both effects are known to reduce transport when acting separately, it has been shown that in some regions of parameter space, noise-induced dephasing induced by their combined effects can counteract localisation and enhance wave transport.
The exploration of waves in disordered media and localisation constitutes a highly intricate field, unveiling hitherto unknown phenomena such as Lévy flight~\cite{Barthelemy2008} and hyper-transport~\cite{Levi2012}. Significantly, the dynamic evolution of disorder introduces nontrivial effects, as illustrated by Levi \textit{et al.} \cite{Levi2012} in a paraxial optical context. Their findings revealed that an evolving random medium induces stochastic acceleration, facilitating the momentum exchange between the wave packet and the disorder. This process transcends diffusion, culminating in the attainment of ballistic transport \cite{Levi2012}. Recently, these features have proven to be useful to improve quantum transport in programmable nanophotonic processors \cite{Harris2017}, enhance energy transport in quantum networks \cite{Maier2019}, and understand fluidization processes in glasses by X-ray irradiation \cite{Dallari2023}. The universality of this phenomenon extends its relevance to diverse wave systems featuring disorder.

Motivated by these observations,  our investigation explores the propagation of elastic waves within a one-dimensional random medium featuring both frozen and evolving disorders in the mass-density distribution (diagonal disorder). Through extensive numerical simulations, we demonstrate that evolving diagonal disorder enhances the wave speed of elastic waves. Our investigation considers random fluctuations following a delta-correlated zero-mean Gaussian distribution. Additionally, we explore the effects of chaotic fluctuations in the mass density of the propagating medium. We show that the system exhibits an equivalent enhanced speed effect of elastic waves under chaotic dynamics. Thus, evolving randomness is not a necessary condition for the increased transport of wavefronts in evolving media, an observation that leads us to coin the term \emph{chaotic speed enhancement of waves in complex media}. Moreover, we show that contrary to the case of the stochastic speed enhancement effect, where transport is enhanced as the noise strength increases, there is a band of optimal levels of chaos where the system reaches a plateau of speed enhancement. Thus, highly evolved chaos does not always give larger speed-up effects.

The remainder of the article is organised as follows. In Section \ref{Section:Models}, we introduce our model for elastic wave propagation in one-dimensional heterogeneous materials with diagonal disorder. Here, we also introduce the models of disorder used for random frozen disorder, stochastic evolving disorder, and chaotic evolving disorder. In Section \ref{Sec:Results}, we show a summary of our results from numerical simulations. We first study in Section \ref{Sec:Random} the effects of complexity through randomness (for frozen and evolving media), and in Section \ref{Sec:Chaos} the effects of introducing complexity through chaos. Finally, we conclude and summarise our main results in Section \ref{Sec:Conclusions}.

\section{Elastic waves in heterogeneous media with diagonal disorder \label{Section:Models}}

\subsection{The model}

We model a random heterogeneous medium that is, in some sense, close to a deterministic homogeneous reference medium~\cite{ostoja2007}. More abstractly, a random (complex) material is represented by a random medium $\mathcal{M}$ member of a family of random media $\mathcal{M}(\zeta)$, where $\zeta$ is a point in the space of events $\Lambda$ (sample space) \cite{ostoja2007, Gardiner2009}. In probability theory, $\Lambda$ consists of all possible outcomes $\zeta$ of an experiment or observation performed on a random system. The statistical description of the system is completed by providing a probability density function over the members of $\mathcal{M}(\zeta)$ \cite{ostoja2007, Gardiner2009}. The system's physical properties are obtained through an ensemble average over multiple observations of the system so that the final results are statistically independent and could be related to experimental results. However, most heterogeneous materials are nonequilibrium, quenched disordered media for which no true statistical ensemble exists. Thus, our description of a heterogeneous system as a random medium corresponds to a heuristic model for the local random spatial variations of physical properties, limiting our study to simple models in which each member of the family $\mathcal{M}(\zeta)$ differs slightly from a homogeneous reference medium $\mathcal{M}^0$. In particular, we will consider random fluctuations introduced in the mass density $\rho$ of the medium, i.e.,
\begin{equation}
\label{Eq:01} \rho(x,t)=\rho_0\left[1+\epsilon\eta(x,t)\right],\quad0\leq\epsilon\ll 1,\quad\vert\eta\vert\leq1,
\end{equation}
where $\eta(x,t)$ is a stochastic process which can also be random in space, $\rho_0$ is the mass density of the homogeneous reference medium, and $\epsilon$ is the complexity strength. Here, we coined the term \textit{random frozen disorder} for the case of a variable $\eta=\eta(x)$ that only depends on space. Alternatively, we shall use the term \textit{random evolving disorder} for the case of a spatiotemporal variable $\eta=\eta(x,t)$. The random variable $\eta$ can only take values from -1 to 1, and the dimensionless small parameter $\epsilon$ can be regarded as a measure of the medium's departure from homogeneity. Equation~\ref{Eq:01} describes what is known as \emph{diagonal disorder} \cite{Ziman1979} since it captures from a microscopic point of view fluctuations of a one-cite parameter: the mass of particles constituting the propagation medium.

Elastic waves propagating in a one-dimensional heterogeneous medium, whose linear mass-density $\rho$ varies randomly in space, are governed by the following hyperbolic wave equation,
\begin{equation}
\label{Eq:02}
  \partial_{tt}\phi-c^2(x,t)\partial_{xx}\phi=0,
\end{equation}
where $\phi$ is the dimensionless perturbation, and $c(x,t):=\sqrt{K_0/\rho(x,t)}$ is the phase velocity of waves. We assume that $K_0$ represents the bulk modulus of the homogeneous reference medium.
In Appendix~\ref{App:1}, we provide a brief derivation of Eq.~(\ref{Eq:02}) from a microscopic model of elastic waves in a one-dimensional system with diagonal disorder. In the same spirit of the models studied in Refs.~\cite{bahraminasab2007, Shahbazi2005, Allaei2006, Allaei2008}, Eq.~(\ref{Eq:02}) can be regarded as a model based on elastic wave propagation in homogeneous media, where we have introduced spatial variations in the phase velocity to account for micro-structural density disorder \cite{Marin2013, Alcantara2018, Discacciati2022, Discacciati2023}.

The phenomenon of Anderson localization, where waves become spatially localized due to disorder, is well-documented in various physical systems \cite{Lagendijk2009, Modugno2010, Segev2013}. For one-dimensional systems with static disorder, it is known that all eigenmodes are typically localized, resulting in the suppression of wave propagation over long distances \cite{Anderson1980, Kramer1993}. This strong tendency for localization arises because, in one-dimensional media, even weak disorder is sufficient to induce exponential localization of the wave modes \cite{Kramer1993, Freilikher1992, Comtet2013}.

In the context of the hyperbolic wave equation studied here, the interplay between static and dynamic disorder leads to distinct propagation dynamics. For static disorder, the localization of wave modes is consistent with the classical theory of Anderson localization, as explored in rigorous analyses of similar wave equations in 1+1 dimensions, including additive and multiplicative random perturbations, such as Refs.~\cite{Figotin1996, Bourgain2004, Balan2019, Balan2022, Balan2022-2} to mention a few. Specifically, we observe that spatial randomness induces localization, which explains why the wave packet does not exhibit unbounded propagation despite the presence of an initial finite velocity.

When the disorder in Eq.~\ref{Eq:02} is time-dependent, however, the dynamics change significantly \cite{Levi2012, Min2013, Sharabi2021, Apffel2022, Carminati2021, Selvestrel2024, Pierrat2024}. In this case, the temporal fluctuations can prevent full localization, allowing for a slight increase in the average wave velocity, which we refer to as {\it enhanced speed effect}. This behaviour highlights the complexity of wave propagation in dynamically disordered systems, where the interplay between spatial and temporal disorder influences the effective transport properties.

\subsection{Models of disorder \label{Sec:ModelsDisorder}}

We consider three types of statistics for the disorder in this work:

\begin{itemize}

\item The first type of disorder assumes that statistical fluctuations of $\eta$ obey a truncated Gaussian-type delta-correlated distribution in space with zero mean \cite{Gardiner2009}, i.e.,
%
\begin{equation}
\label{Eq:03a}
\langle \eta(x,t)\rangle = 0,
\end{equation}
\begin{equation}
\label{Eq:03b}
C(x,t):=\left\langle\eta(x',t')\,\eta(x+x', t+t')\right\rangle_{x',t'}=\sigma^2 \delta(x)\delta(t),
\end{equation}
%
where $\sigma$, $\delta$, and $\langle\cdot\rangle$ denote the variance, the Dirac delta function, and the ensemble average with respect to the underlying Gaussian probability distribution, respectively. This type of white noise is a continuous stochastic process in space and time and thus admits the application of the standard theory of stochastic differential equations \cite{Gardiner2009, Oksendal2020}. In our numerical simulations, we employed the Ziggurat method~\cite{Marsaglia1984} to generate statistics of this kind, which is a widely employed method to model various noisy sources in nature \cite{Gardiner2009}. Notice that from Eqs.~\ref{Eq:01},~\ref{Eq:03a} and~\ref{Eq:03b} we have $\langle \rho\rangle=\rho_0$. We used a truncated Gaussian as the underlying probability distribution for $\eta$ to ensure that $\vert\eta\vert\leq1$, thus keeping the mass density in Eq.~\ref{Eq:01} always positive even for relatively large values of the variance.

\item We introduce a \juan{second} type of evolving disorder modelled from a complex but deterministic source. In particular, at $t=0$, we considered a random distribution of uncorrelated values uniformly distributed in space for the field $\eta(x,t=0)$. Then, for each coordinate $x=x^*$, we introduced chaotic evolving disorder through time series generated from the well-known logistic map \cite{Strogatz2018}\footnote{It is important to remark that there is no coupling between spatial nodes in the spatiotemporal dynamics of $\eta$, and thus the chaotic evolution is independent of the spatial disorder.},
\begin{equation}
\label{Eq:04}
\eta_{n+1}(x^*)=r\eta_{n}(x^*)\left[1-\eta_{n}(x^*)\right],
\end{equation}
famously known for exhibiting intermittency and a period-doubling route to chaos as the bifurcation parameter $r$ increases within the interval $[0,4]$. Evolving chaos is obtained for $r>r_c\simeq3.56995$, except for the isolated ranges corresponding to the periodic windows of the logistic map. One obtains completely developed chaos for $r=4$.

\item We consider a third type of deterministic evolving disorder extracted from chaotic fluctuations of the following continuous-time three-dimensional dynamical system:
\begin{equation}
    \label{Eq:Lorenz}
    \frac{\mbox{d}x_1}{\mbox{d}\tau}=\alpha(x_2-x_1),\quad\frac{\mbox{d}x_2}{\mbox{d}\tau}=x_1(\mu-x_3)-x_2,\quad\frac{\mbox{d}x_3}{\mbox{d}\tau}=x_1x_2-\beta x_3,
\end{equation}
which is the renowned Lorenz model \cite{Lorenz1963, Ott2002, Lynch2004, Strogatz2018}. We used $\mu$ as our tunable bifurcation parameter for the Lorenz model~\ref{Eq:Lorenz} to pass from non-chaotic to chaotic behaviour, keeping parameters $\alpha=10$ and $\beta=8/3$ fixed. As before, we assumed a random distribution of uncorrelated values at $t=0$, uniformly spread across space for the field $\eta(x,t=0)$ in Eq.~\ref{Eq:01}. Subsequently, we introduced time-dependent chaotic disorder for each spatial coordinate $x=x^*$ by generating a series from the signal $x_i(t)$ $(i=1,2,3)$ with the largest Lyapunov exponent.

\end{itemize}

We have introduced a different dimensionless time $\tau$ in the continuous nonlinear dynamical system~\ref{Eq:Lorenz} to differentiate the proper time scale of the chaotic dynamics from the time scale of the wave equation~\ref{Eq:02}. Indeed, many experimental sources of chaos could be employed in laboratories to artificially feed the complexity in Eq.~\ref{Eq:01}, including nonlinear integrated circuits, mechanically coupled oscillators, and arrays of laser diodes in nonlinear optical resonators \cite{Ott2002}. For instance, elastic waves can have a period of a few seconds in some solids \cite{Achenbach2012}, whereas chaotic fluctuations in nonlinear optical resonators can have a characteristic time of the order of microseconds \cite{Lugiato2015}. Thus, the characteristic time $\tau$ of the chaotic system can be entirely different to the characteristic time of the elastic wave dynamics. For our numerical simulations, we considered $t/\tau=10$ so that the chaotic dynamics of $\rho(x,t)$ in Eq.~\ref{Eq:01} is sufficiently fast and irregular to induce nontrivial effects, such as the speed-enhancement effect reported in this work.

Notice that the dimensionless parameter $\epsilon$ in Eq.~\ref{Eq:01} represents the noise strength in the case of stochastic disorder. In contrast, for the deterministic case where $\eta$ is a chaotic signal, $\epsilon$ can be interpreted as the strength of chaos. Thus, while $\epsilon$ quantifies the amplitude of fluctuations in both cases, the nature of these fluctuations—random or chaotic—leads to distinct system dynamics.

\section{Results \label{Sec:Results}}

We simulate numerically the time evolution of an initially-at-rest Gaussian wave packet of width $\sigma_p$, given by
\begin{equation}
\label{Eq:05}
\phi(x,t=0)=\exp\left(-\frac{(x-x_0)^2}{2\sigma_p^2}\right),
\end{equation}
governed by Eq.~\ref{Eq:02} considering free-end (Neumann) boundary conditions. Hereon, we shall consider only dimensionless variables where the space $x$ is measured in units of $\sigma_p$, and time $t$ is measured in units of $\sigma_p/c_0$, where $c_0=\sqrt{K_0/\rho_0}$ is the phase speed of elastic waves in the homogeneous reference medium. At $t=0$, the wavepacket is stationary and centred at $x_0=L/2$ (the centre of the space domain), with $L=60$. We used 800 discrete nodes in space and finite differences of second-order accuracy for the space derivatives with step $\Delta x=0.075$. For the time integration, we used the Dormand-Prince algorithm as an explicit Runge-Kutta of orders 4 and 5 with an adaptive time step \cite{MatlabODE45}. We have ensured that the Courant-Friedrichs-Lewy (CFL) condition is fulfilled at each realisation of the complex fluctuations in the phase speed. This guarantees that the observed asymmetry in backscattering patterns, as we will see later, is not a result of numerical artefacts but rather stems from the intrinsic randomness of the spatial disorder. For each statistical realisation of the disorder, we have computed the speed of the counter-propagating Gaussian pulses through the slope of the linear interpolation of its position as a function of time, obtained directly by tracking the wavefront position as a function of time. We calculated 500 statistical realisations of the disorder for each combination of parameters using a parallel code written in \textsc{Matlab} \copyright, and we have verified that the ensemble average of the speed and the amplitude fields becomes statistically invariant.

\begin{figure}
    \centering
    \scalebox{0.6}{\includegraphics{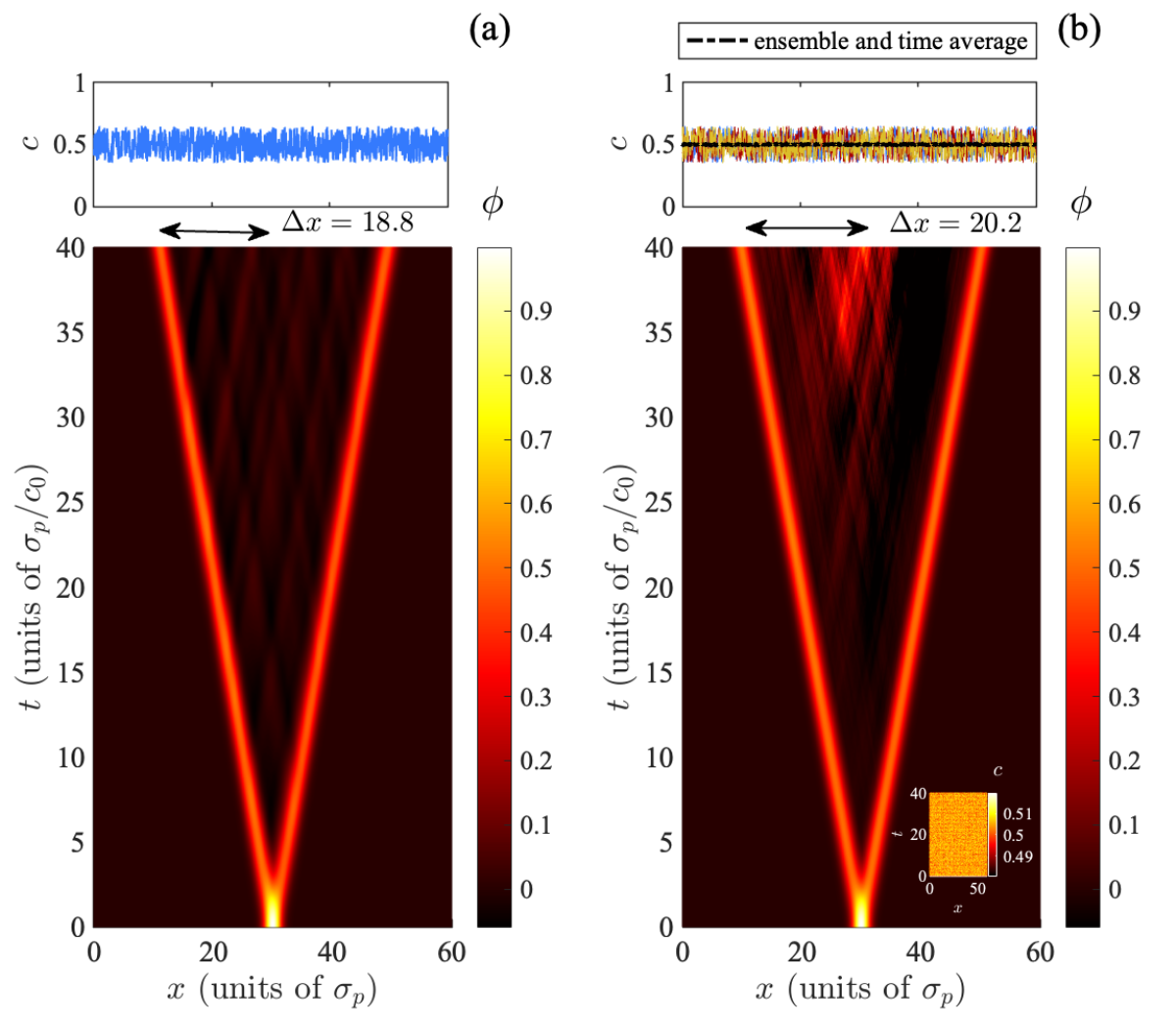}}
    \caption{Typical behaviour of waves in random complex media with an underlying delta-correlated Gaussian probability distribution. Cases are shown for a single disorder realisation with disorder strength $\epsilon=0.3$. The colour scale indicates the field of the elastic wave $\phi$. \textbf{(a)} Evolution under frozen randomness. The upper panel shows the random fluctuations of the phase velocity $c$ in space for a single realisation. \textbf{(b)} Evolution under evolving randomness. The inset shows a realisation of the random disorder in the phase speed. The upper panel shows the random fluctuations of the phase velocity $c$ in space at $t=0$ for various disorder realisations, along with the time and ensemble-averaged speed over 500 realisations (black dashed-dotted line).}
    \label{Fig:01}
\end{figure}

\subsection{Complexity through randomness: frozen and evolving media \label{Sec:Random}}

Figure~\ref{Fig:01} shows typical instances of wave propagation in a random media obtained under frozen randomness [Fig.~\ref{Fig:01}(a)] and evolving randomness [Fig.~\ref{Fig:01}(b)]. We considered delta-correlated Gaussian distributions with $\epsilon=0.3$ in both cases. In the trivial homogeneous case where the phase speed $c$ is constant, it is well known that under Eq.~\ref{Eq:02}, an initially at-rest Gaussian wavepacket breaks into two counterpropagating pulses with half the amplitude of the initial pulse. In Figure~\ref{Fig:01}(a), the initially at-rest Gaussian wavepacket also breaks into two counterpropagating pulses in our heterogeneous material. These two pulses propagate at nearly constant speed, and multiple backscattering due to the spatial disorder within the density of the medium produces a small amplitude interference pattern observed in the region between them. The upper panel in Fig.~\ref{Fig:01}(a) shows a typical statistical realisation of the frozen disorder induced in the phase speed in the heterogeneous medium, which is indeed random in space with mean value $c_0=0.5$.

For the case of evolving randomness, Fig.~\ref{Fig:01}(b) shows that the multiple backscattering produces a larger amplitude interference pattern than the case of frozen randomness. Notice that the backscattering patterns are not symmetric with respect to the point $x=x_0$ for frozen and evolving disorders. This asymmetry arises because each realization of the spatial disorder is generally not symmetric. However, the patterns under evolving disorder are more complex and emerge from the spatiotemporal complexity of the phase velocity (see inset in Fig.~\ref{Fig:01}(b)). The upper panel in Fig.~\ref{Fig:01}(b) shows several statistical realisations of the phase velocity at $t=0$. The ensemble average of the phase speed after 500 disorder realisations gives an almost constant phase speed $\langle c\rangle=c_0=0.5$, as shown with a black dashed-dotted line in the upper panel of Fig.~\ref{Fig:01}(b).

\begin{figure}
    \centering
    \scalebox{0.35}{\includegraphics{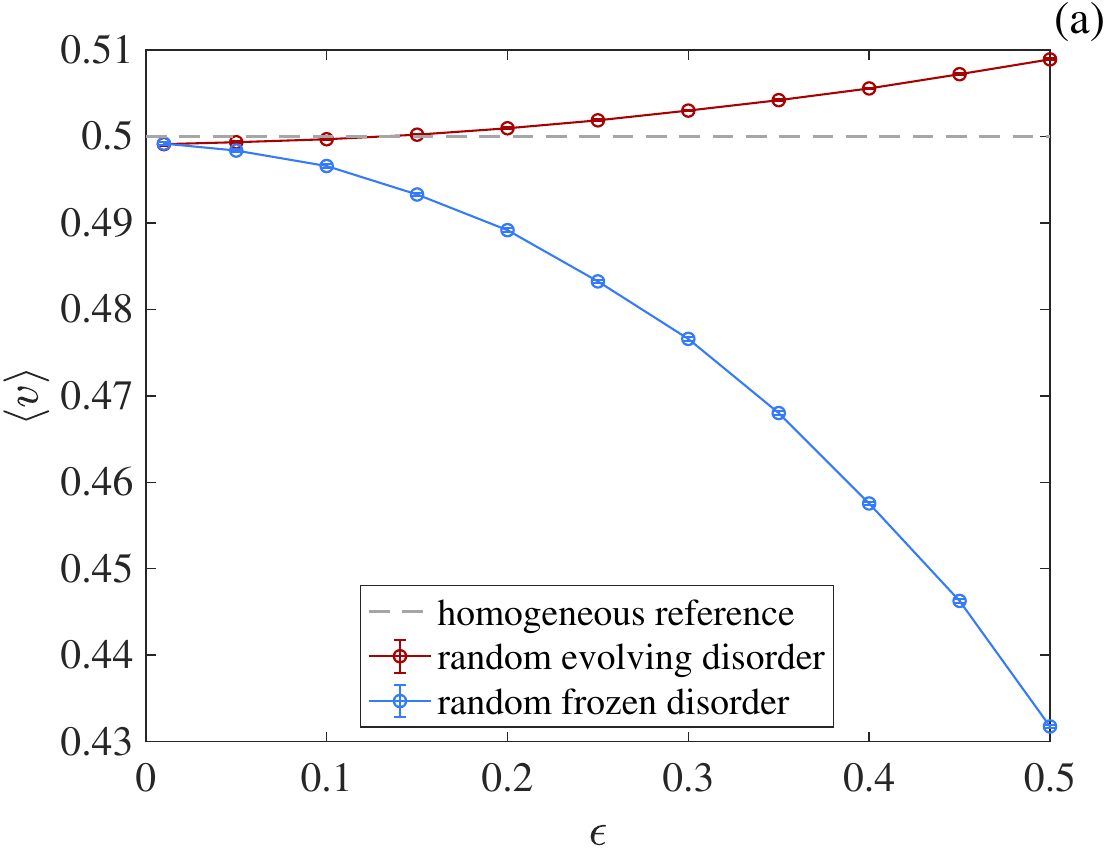}\includegraphics{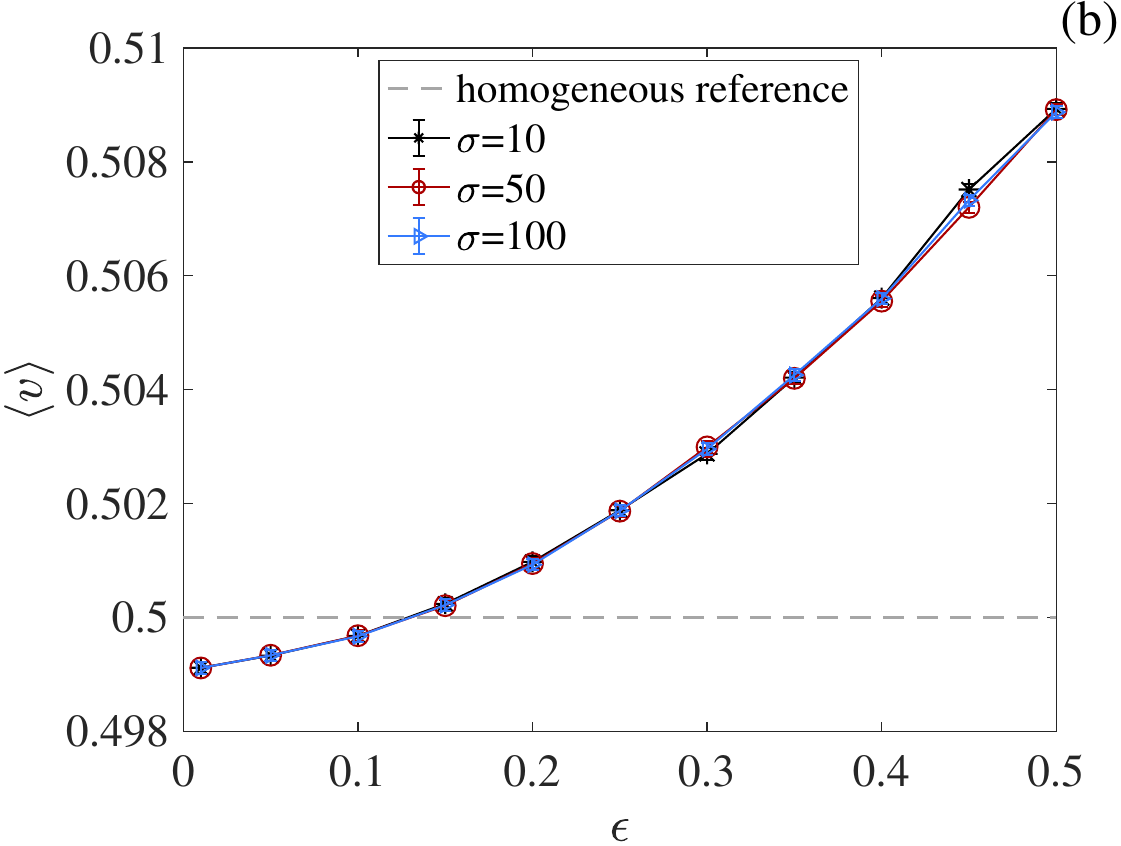}}
    \scalebox{0.5}{\includegraphics{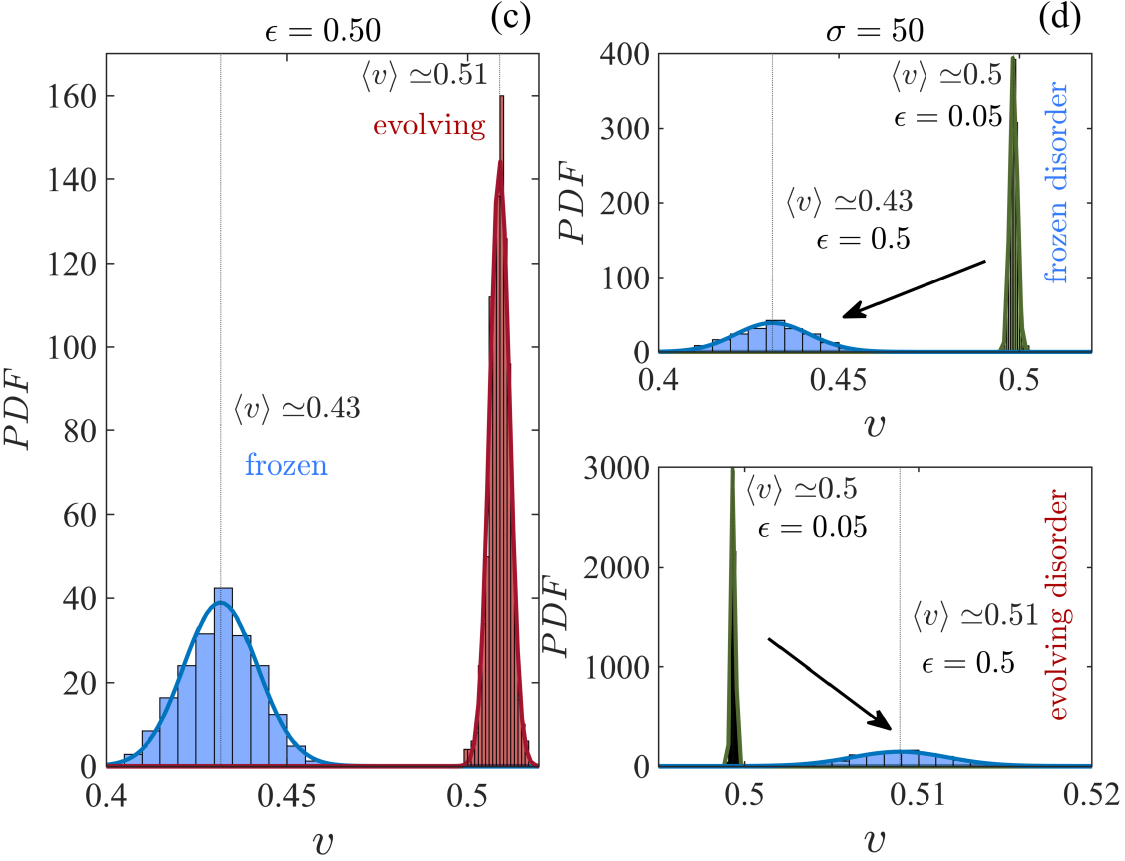}}    
    \caption{Stochastic-induced speed enhancement of elastic waves: Ensemble average speed $\langle v\rangle$ as a function of the disorder strength $\epsilon$ for delta-correlated normal distribution with zero mean. \textbf{(a)} Comparison of the measured values of $\langle v\rangle$ in numerical simulations under evolving and frozen random disorder with variance $\sigma=50$. For random frozen (evolving) disorder, the speed decreases (increases) for increasing values of $\epsilon$.  \textbf{(b)} Dependence of the ensemble average speed on the variance of the white noise (evolving disorder). The stochastic-induced speed enhancement of the wave under evolving randomness remains robust, appearing with no qualitative differences under changes in the variance. Error bars are too small to be distinguishable in the scale. \textbf{(c)} Probability distribution functions (PDF) of the velocity $v$ for $\epsilon=0.5$ and $\sigma=50$. The mean of the PDF is shifted towards a higher velocity when the disorder passes from frozen to evolving. \textbf{(d)} Dependence of the PDF on the disorder strength $\epsilon$. Arrows indicate the transition when increasing the value of $\epsilon$.}
    \label{Fig:02}
\end{figure}

We have noticed that the counterpropagating pulses travel faster under random evolving disorder than under random frozen disorder. As indicated with double arrows in Fig.~\ref{Fig:01}, during the same time interval $\Delta t=40$, the pulse travels a distance $\Delta x\simeq18.8$ under random frozen disorder and $\Delta x\simeq20.2$ under random evolving disorder. This observation suggests a speed enhancement process occurs in the evolving medium compared to the frozen case. To inquire about this observation, we computed for different values of $\epsilon$ the ensemble average of the pulse speed, $\langle v\rangle$, and compared the observed behaviour between frozen and evolving cases. Figure~\ref{Fig:02}(a) reveals that as the strength of frozen randomness increases, the speed of travelling pulses decreases, and thus the wave tends to localise. Indeed, spatial randomness is expected to induce a strong tendency for waves to localise in one-dimensional systems \cite{Kramer1993, Freilikher1992, Comtet2013}. Surprisingly, Fig.~\ref{Fig:02}(a) also reveals that the speed slightly increases as the strength of the evolving randomness increases.  Moreover, for $\epsilon$ sufficiently high, the mean speed of the wavefront surpasses the reference value corresponding to the homogeneous medium, \textit{i.e.} $c_0<\langle v\rangle$. Thus, increasing the strength of disorder in evolving media, rather than promoting wave localisation, enhances wave transport. Consequently, for a fixed value of $\epsilon$, waves travel faster under a random evolving disorder than under a random frozen disorder.

Thus, we have discovered a stochastic-induced speed enhancement in the propagation of elastic waves. This phenomenon is similar to that observed by Levi \textit{et al.} in Ref.~\cite{Levi2012}, where hyper-transport of light by evolving disorder is reported numerically and experimentally using an effective 2+1-dimensional paraxial optical setting. Here, we report a similar phenomenon for elastic waves in an evolving 1+1-dimensional randomly disordered medium. However, instead of observing an accelerated wave with a power-law dependence of the form $\langle v\rangle\propto t^{\gamma}$ with $\gamma>0$, as observed in Ref.~\cite{Levi2012}, here we have obtained a wave with speed enhanced by the disorder following a law $\langle v\rangle\propto f(\epsilon)$, with $f$ an increasing function of $\epsilon$.

Following the previous observations, it is natural to wonder about the effect of a truncated Gaussian probability distribution with a higher dispersion of sample points. Thus, we repeated our analysis with a variance $\sigma=100$, i.e. twice as that in Fig.~\ref{Fig:02}(a). Figure~\ref{Fig:02}(b) shows almost no differences in the outcome from the ensemble average speed. Samples of a random variable under a Gaussian probability distribution predominantly accumulate near its mean value, and extreme events corresponding to the tails of the truncated Gaussian distribution are less probable to occur. Thus, although extreme events are more probable for $\sigma=100$ than for $\sigma=50$, they are still much less probable than events near the mean of the stochastic process. Thus, since our underlying probability distribution is truncated, the effect of enhanced transport under evolving randomness is qualitatively the same for both variance values. The observed phenomenon is robust against changes in the variance.

We also considered the case $\sigma=10$ in Fig.~\ref{Fig:02}(b) and obtained the same behaviour: the evolving randomness counteracts wave localisation despite the small value of the variance. However, we expect the behaviour to change if the variance is too small. Since $\vert\eta\vert\leq1$, we expect the speed-enhancement effect to disappear for $\sigma\ll1$ as $\sigma\to0$.

To gain a deeper insight into the statistical properties of the speed-enhancement effects, we show in figures~\ref{Fig:02}(c-d) the probability distribution function (PDF) for different values of $\epsilon$ under frozen and evolving randomness.  PDFs have been computed from the fit of a beta distribution function to the histograms of the wavefront speed. The beta distribution provides greater flexibility than the normal distribution and is usually applied to model the behaviour of random variables limited to finite-length intervals ~\cite{Papoulis2002}. In Fig.~\ref{Fig:02}(c), we show that for the same parameter values, simply changing the randomness from frozen to evolving induces a shift of the PDF towards a sharper distribution with a higher mean speed. In Fig.~\ref{Fig:02}(d), we show that increasing the value of $\epsilon$ always has the effect of increasing the variance of the wave speed: no matter if the disorder is frozen or evolving, larger disorder strength leads to wider PDFs. However, increasing the value of $\epsilon$ has contrary effects on the mean speed under frozen and evolving disorder: Under frozen (evolving) disorder, increasing the disorder's strength decreases (increases) the mean speed velocity.

\subsection{Complexity through chaos: chaotic speed enhancement in complex media \label{Sec:Chaos}}

Following our findings above, we wondered if the evolving nature of the media must be strictly random to induce the wave speed enhancement of elastic waves. Thus, we explored the outcome of the system if the medium evolves in a deterministic but complex way, following the fluctuations from chaotic signals. We have discovered that chaos also induces the same qualitative speed enhancement effect. Our explorations lead us to coin the term \emph{chaotic speed enhancement of waves in complex media}.

\subsubsection{Speed enhancement under chaos from the logistic map.\label{Sec:Logistic}}

\begin{figure}
    \centering
    \scalebox{0.38}{\includegraphics{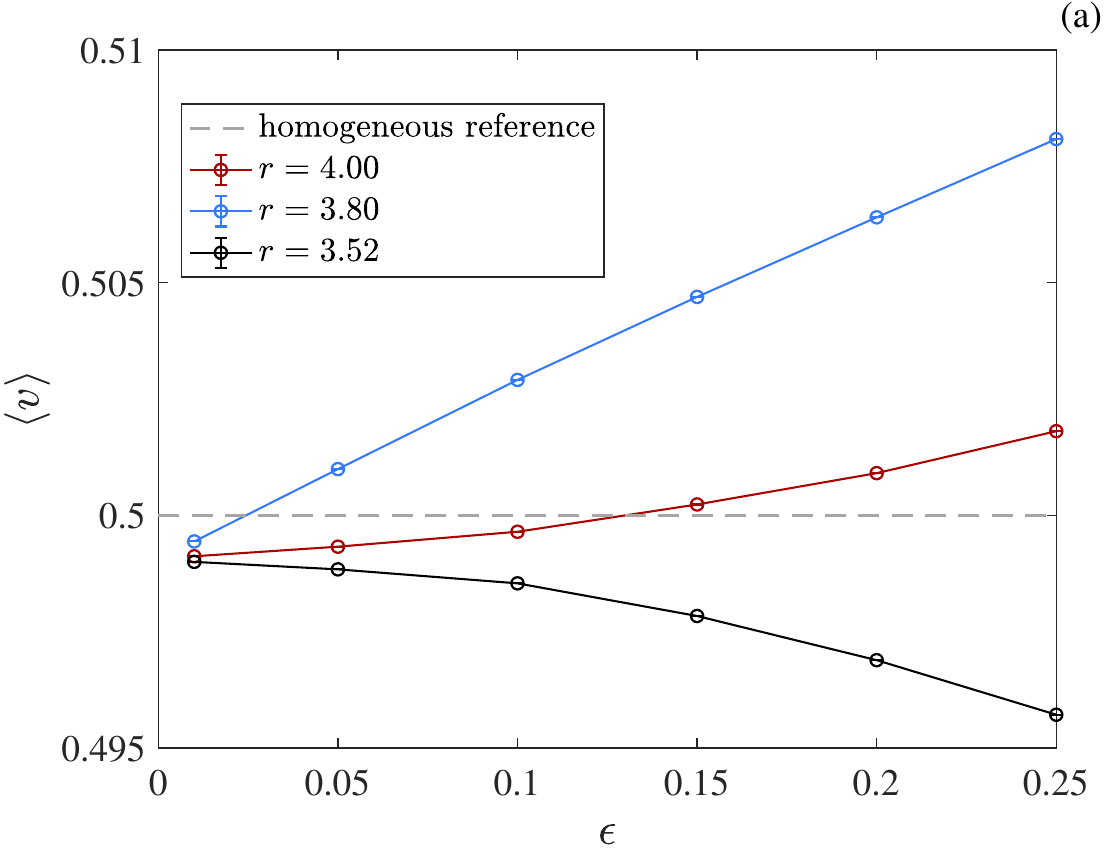}\includegraphics{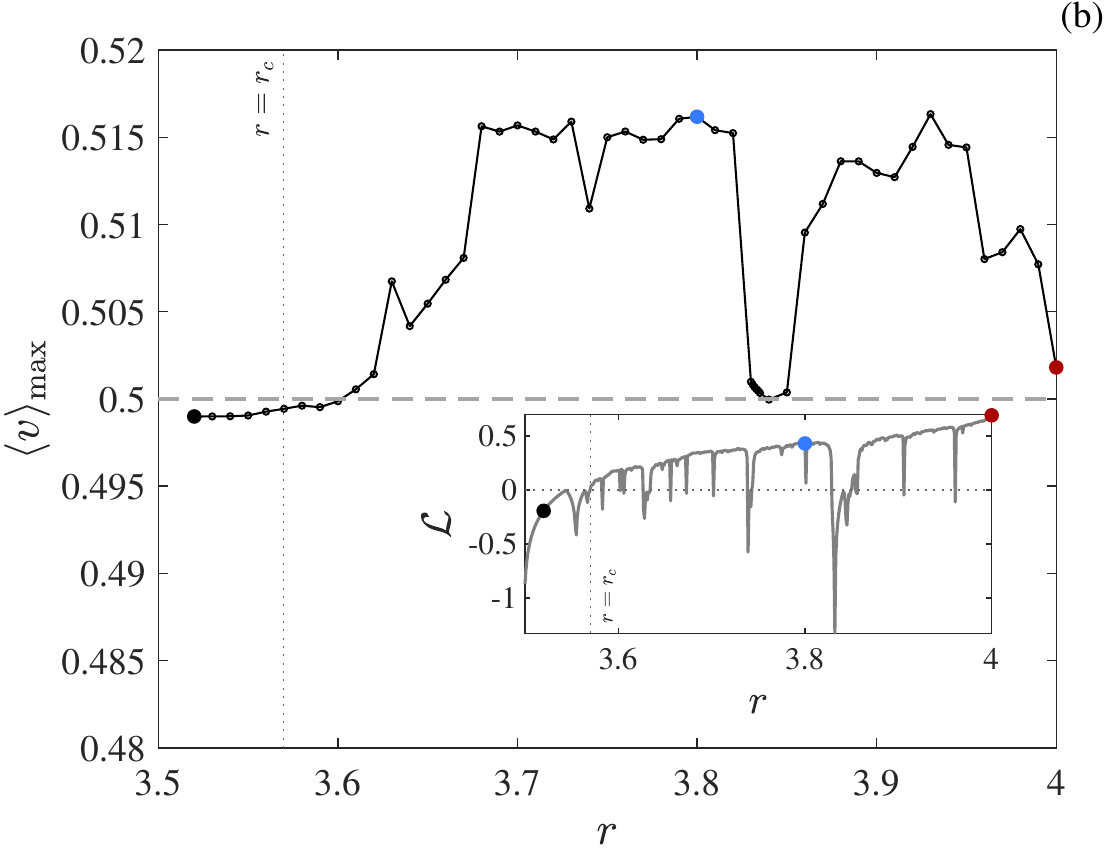}}
    \scalebox{0.4}{\includegraphics{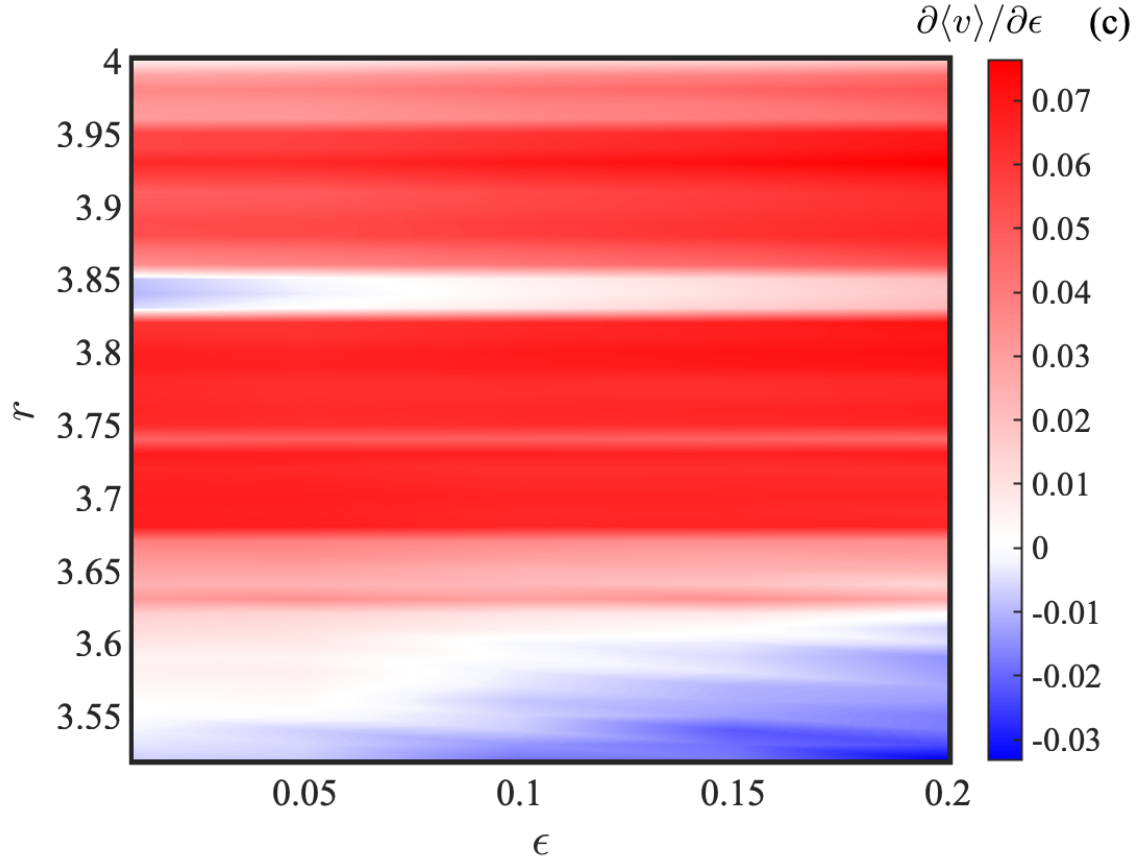}\includegraphics{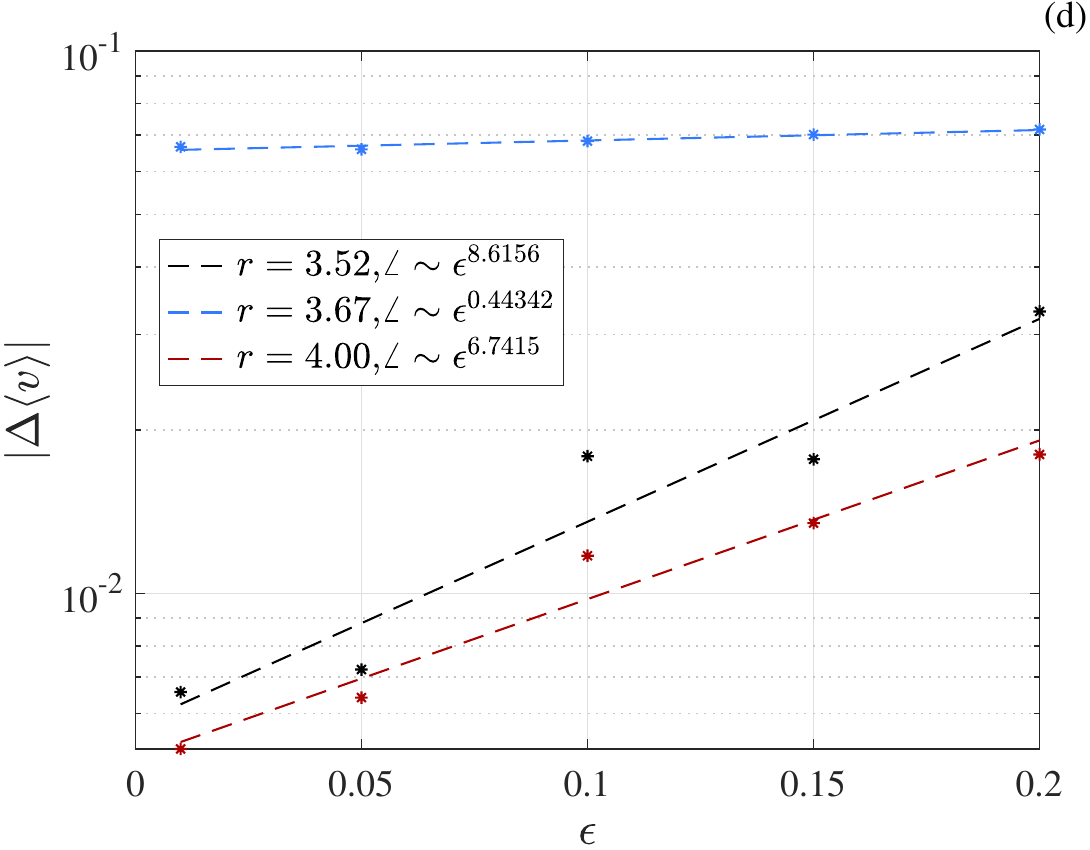}}
    \caption{Chaos-induced speed enhancement of elastic waves: \textbf{(a)} Ensemble average speed $\langle v\rangle$ as a function of disorder strength $\epsilon$ for different values of the bifurcation parameter $r$. \textbf{(b)} Maximum speed of elastic waves as a function of the bifurcation parameter. The maximum values are obtained within a broad bandwidth of $r$. The inset shows the Lyapunov exponent $\mathcal{L}$ as a function of $r$ along with three sample points used to characterise the speed dependence on the chaos level (black, blue and red dots). \textbf{(c)} Phase space representation, indicating regions in the parameter space $(\epsilon, r)$ where localisation (blue regions) and chaos-induced speed enhancement (red regions) are observed. \textbf{(d)} Measured scaling laws of the chaos-induced speed enhancement with the disorder strength.  Error bars are too small to be distinguishable in the scale.}
    \label{Fig:03}
\end{figure}

In Fig.~\ref{Fig:03}(a), we show the ensemble-averaged speed of elastic waves as a function of the disorder strength $\epsilon$ for different bifurcation parameter $r$ values. For $r=3.52$, the logistic map exhibits a periodic orbit with period $4$. We observe that the wave tends to localise as $\epsilon$ increases. In this case, the elastic waves undergo multiple scattering at localised heterogeneities periodically modulated in time. Thus, the wave tends to localise as the amplitude of such modulations becomes larger. 

However, we notice that for $r=3.80$, a value of the bifurcation parameter where the system evolves chaotically, there is a remarkable increase of the wave speed as the disorder strength $\epsilon$ increases, as shown in Fig.~\ref{Fig:03}(a). Indeed, $\langle v\rangle$ increases monotonically with $\epsilon$. Importantly, comparing Fig.~\ref{Fig:02}(a) and Fig.~\ref{Fig:02}(b) with Fig.~\ref{Fig:03}(a), we notice that chaotic dynamics in the evolving medium induces a speed increase of the same order as in the stochastic case.

We have systematically studied the ensemble-averaged speed of waves for increasing bifurcation parameter values. For $r=4$, the value at which the medium exhibits completely evolved chaos, one might expect that the effect of increased transport will reach its stronger regime. However, we have found that the speed-enhancement effect is more moderate than for $r=3.80$, as shown in Fig.~\ref{Fig:03}(a). The latter evidences an optimal value of $r$ (and thus, an optimal level of chaos), where the chaotically induced transport is stronger. 

We have computed the maximum value of $\langle v\rangle$, $\langle v\rangle_{\max}$, in terms of the bifurcation parameter $r$ in Fig.~\ref{Fig:03}(b). We observe that the maximum ensemble-averaged speed reaches a maximum plateau for intermediate values of $r$ within the interval $[r_c,\,4]$, where $r_c\simeq3.56995$ is the critical value for the onset of chaos in the logistic map. A black point in Fig.~\ref{Fig:03}(b) indicates the case $r=3.52$, which corresponds to the non-chaotic dynamics of the system. The blue point in Fig.~\ref{Fig:03}(b) indicates the case where we have detected the largest value of $\langle v\rangle$, corresponding to $r=3.80$. The red point Fig.~\ref{Fig:03}(b) corresponds to the case of completely evolved chaos for $r=4$.

Notice in Fig.~\ref{Fig:03}(b) that there are ranges for the breakdown of the chaotic speed-up effect in the middle of the chaotic speed-up plateau. These breakdowns are associated with the crisis bifurcation of the underlying chaotic system, where the strange attractor of the logistic map suddenly disappears \cite{Ott2002}.  These breakdown ranges of $r$ correspond to sub-ranges of $[r_c,\,4]$ where the speed decays abruptly due to the emergence of the periodic windows of the logistic map. Indeed, in the inset of Fig.~\ref{Fig:03}(b), we show the Lyapunov exponent $\mathcal{L}$ of the logistic map as a function of $r$ (the methods used to estimate $\mathcal{L}$ are shown in Appendix~\ref{App:2}), evidencing that the broadest range of speed-up breakdown corresponds to the wider periodic window of the logistic map. For each periodic window of the logistic map, which is fractally distributed within the range $[r_c,\,4]$, there will be a corresponding breakdown of the chaotic speed-up effect. Also, in the inset of Fig.~\ref{Fig:03}(b), we show the cases corresponding to $r=3.52$, $r=3.80$ and $r=4$, with black, blue, and red dots, respectively, to emphasise that highly evolved chaos do not imply higher speed-up effects: an optimum value is somewhere in between the points $r=r_c$ and $r=4$.

\begin{figure}
    \centering
    \scalebox{0.5}{\includegraphics{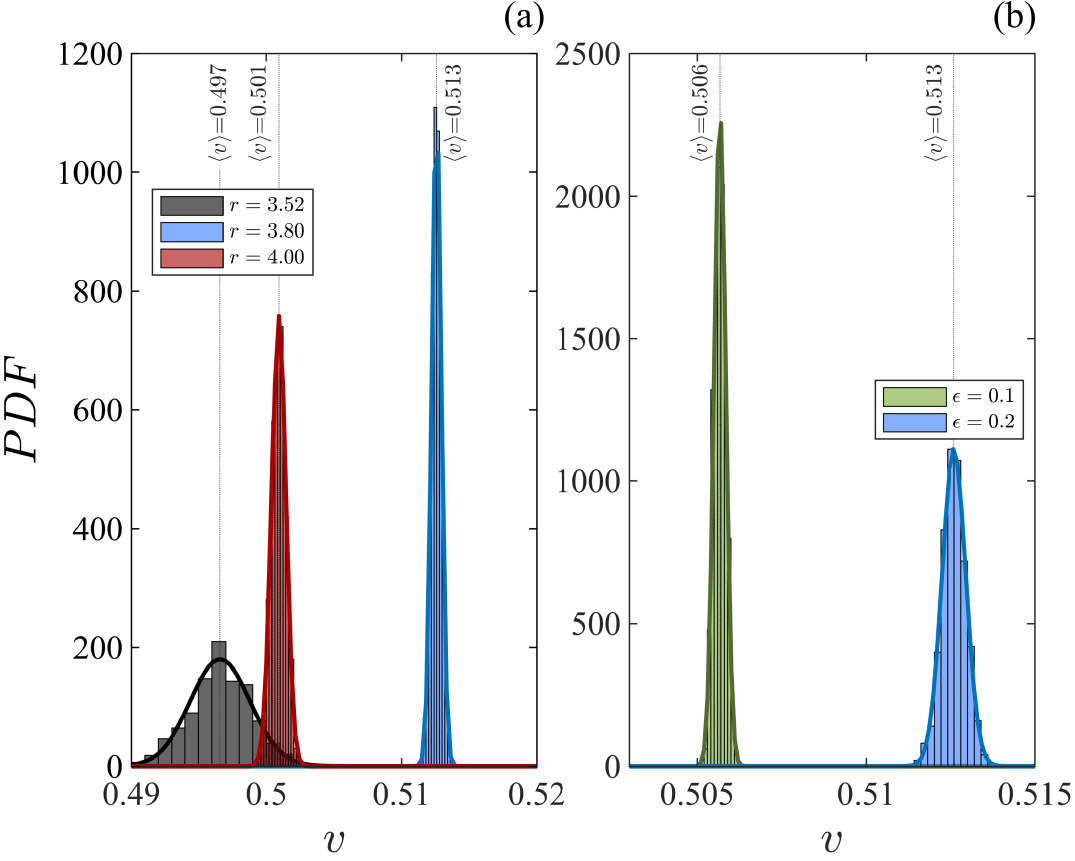}}
    \scalebox{0.4}{\includegraphics{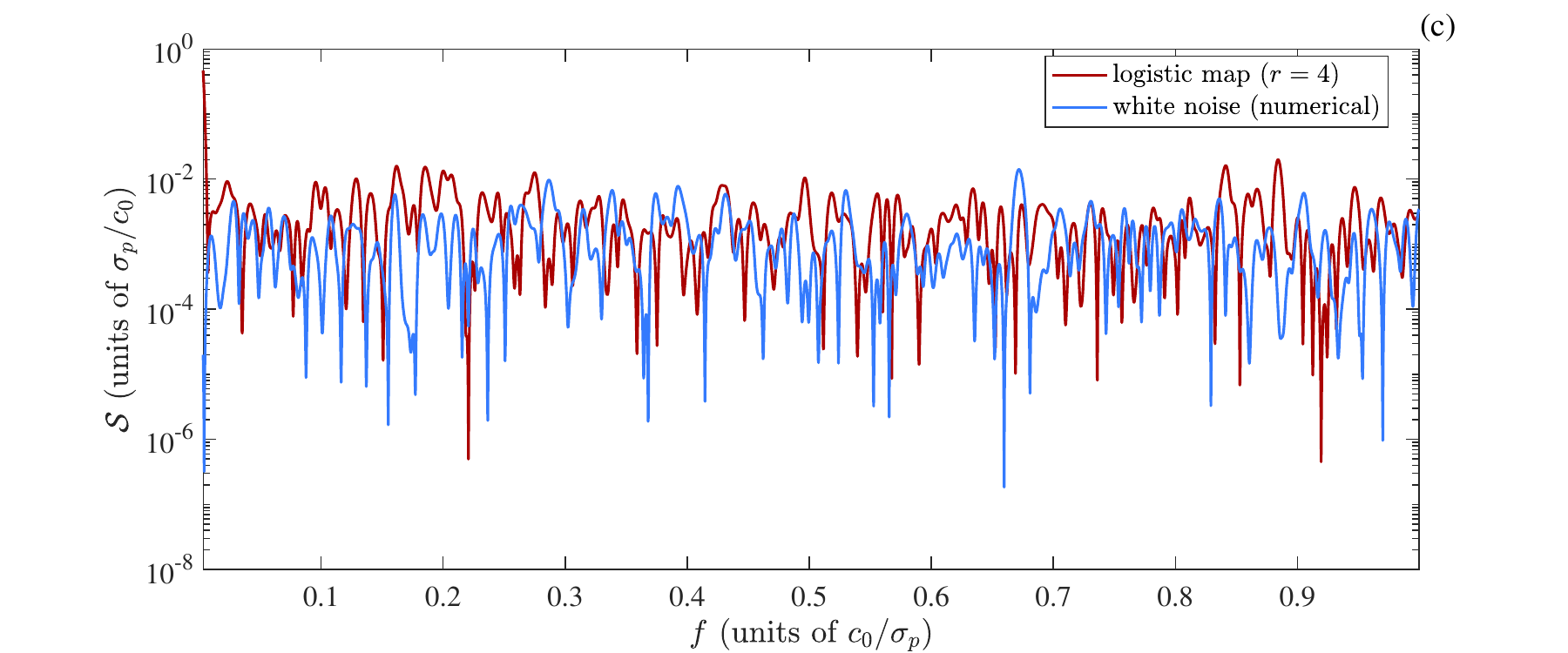}}
    \caption{Statistics of the chaotic speed enhancement under the logistic map. \textbf{(a)} Probability density functions (PDF) of the speed $v$ for fixed chaos strength ($\epsilon=0.2$) and different values of the bifurcation parameter $r$.  \textbf{(b)} Behaviour of the PDFs for different values of $\epsilon$ and $r=3.80$. \textbf{(c)} Comparison of the power spectra $\mathcal{S}(f)$ between numerically generated white noise (with zero mean and unit variance) and the logistic map for $r=4$. }
    \label{Fig:04}
\end{figure}

Figure~\ref{Fig:03}(c) shows a diagram in the parameter space $(\epsilon, r)$ of the chaotic system. Here, the colour scale indicates the value of $\partial\langle v\rangle/\partial\epsilon$. Thus, red regions indicate combinations of parameters for chaotic speed-up, whereas white and blue regions are for localisation. Notice that the breakdowns of chaotic speed-up emerge as white stripes within red regions in parameter space. 

In Fig.~\ref{Fig:03}(d), we study the scaling laws of the ensemble-averaged speed jumps, $\vert\Delta \langle v\rangle\vert$, with the disorder strength $\epsilon$. We observe that, although $\vert\Delta \langle v\rangle\vert$ can reach larger values for $r=3.67$ than for $r=4$, it grows faster with $\epsilon$ for completely evolved chaos ($r=4$) than for moderate chaos ($r=3.67$). Thus, although the largest ensemble-averaged speed occurs for moderate chaos, the ensemble-averaged speed increases faster with $\epsilon$ for completely evolved chaos.

Figure~\ref{Fig:04} summarises the statistical properties of the chaotic speed enhancement effect under the logistic map. Figure~\ref{Fig:04}(a) shows the PDFs of the wavefront speed for three different values of $r$ at a fixed $\epsilon=0.2$. The behaviour of the mean speed is consistent with our analysis in Fig.~\ref{Fig:03}. In Fig.~\ref{Fig:04}(b), we observe that for increasing values of $\epsilon$ ($r=3.80$), increasing the disorder strength leads to an increase in the mean speed $\langle v\rangle$ and the variance, in concordance with the qualitative behaviour observed in Section~\ref{Sec:Random} regarding the variance under evolving randomness.

The similarity between the power spectrum of the chaotic logistic map \juan{for $r=4$} and that of white noise (see Fig.~\ref{Fig:04}(c)) suggests that, in this fully chaotic regime, the system behaves similarly to white noise in a statistical sense. However, it is important to emphasize that, unlike true white noise, which is inherently random, the output of the logistic map remains deterministic. This distinction highlights that while chaotic systems can exhibit noise-like features, they do so through deterministic processes.

This analysis demonstrates how the chaotic dynamics of the logistic map, especially near \( r = 4 \), can mimic stochastic behaviour. The nearly flat power spectrum thus serves as evidence of the system's high complexity and unpredictability, reinforcing its utility as a model for studying chaotic behaviour.

\subsubsection{The case of continuous-time chaotic processes.}

Chaotic signals generated from the logistic map~\ref{Eq:04} do not have a well-defined continuous limit (the one-dimensional continuous logistic model, $\mbox{d}\eta/\mbox{d}\tau=r\eta(1-\eta)$, cannot exhibit chaos). Thus, we wondered if the observed chaotic speed enhancement effect is robust if we considered a continuous-time chaotic process. To answer this question, we have considered time signals generated from the Lorenz system~\ref{Eq:Lorenz}, a paradigmatic model of continuous-time chaotic dynamics.

\begin{figure}
    \centering
    \scalebox{0.4}{\includegraphics{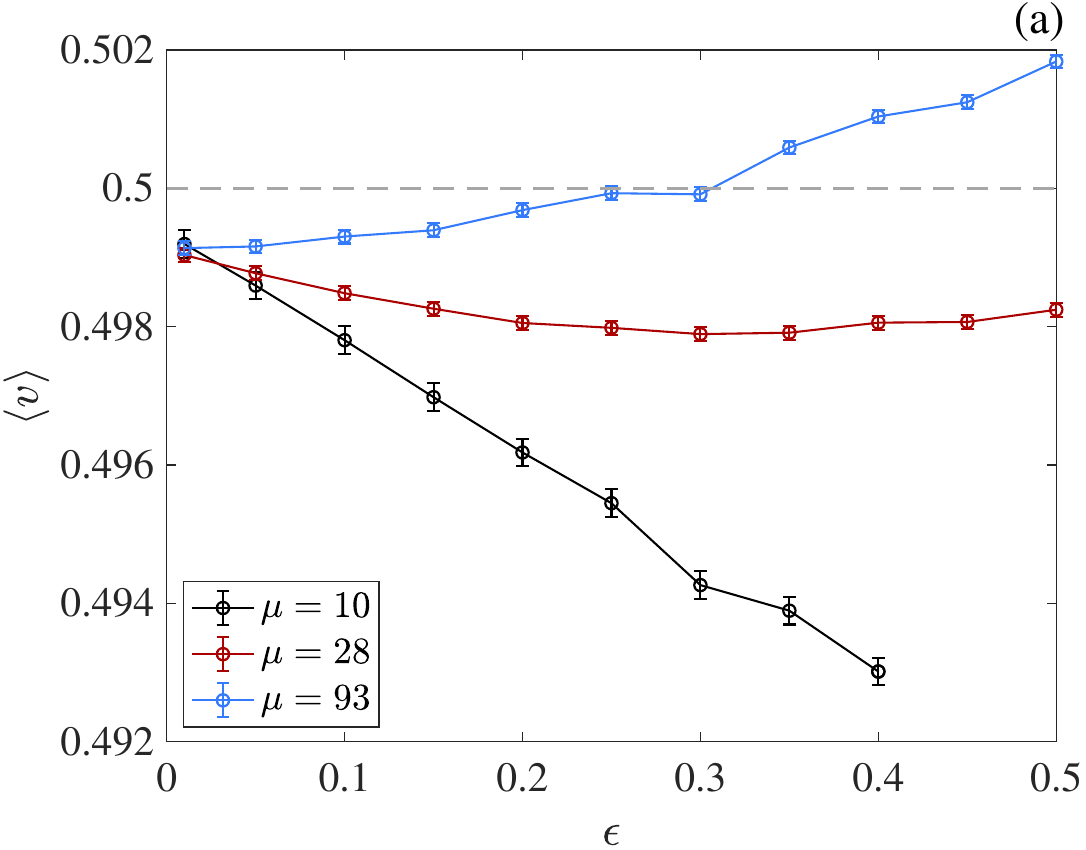}}
    \scalebox{0.4}{\includegraphics{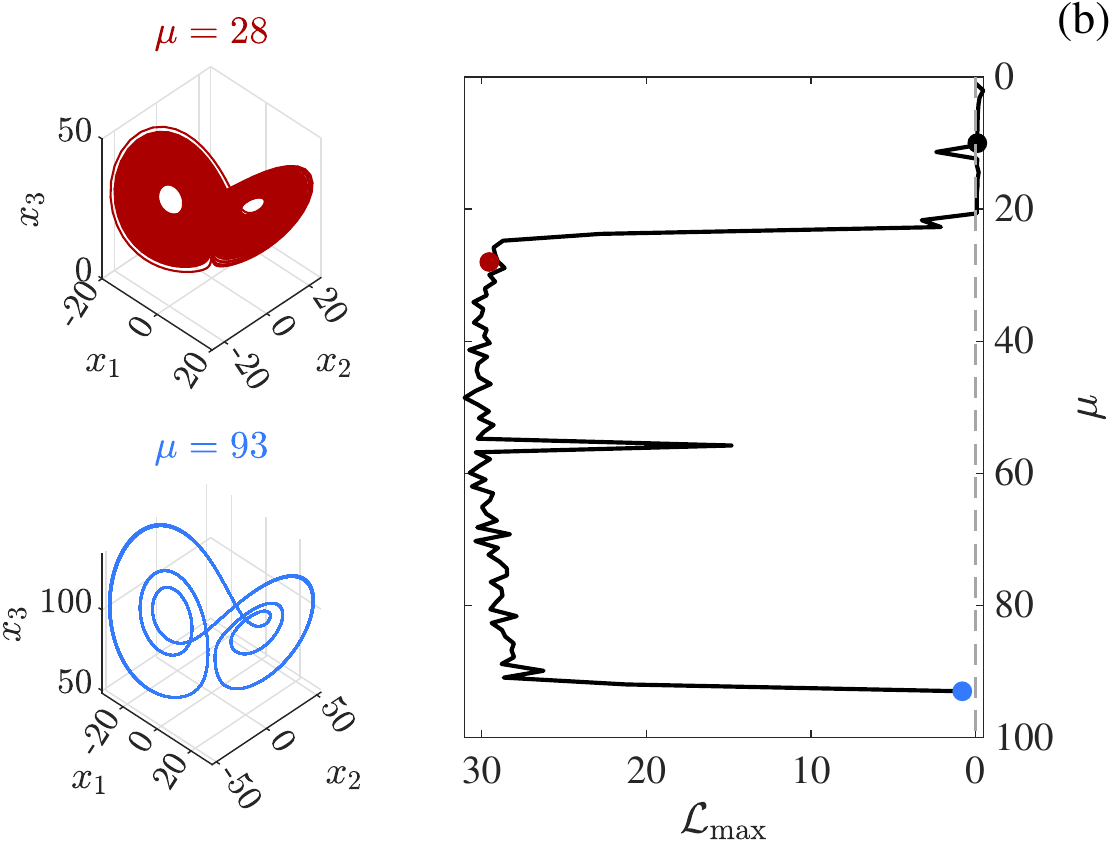}}
    \scalebox{0.4}{\includegraphics{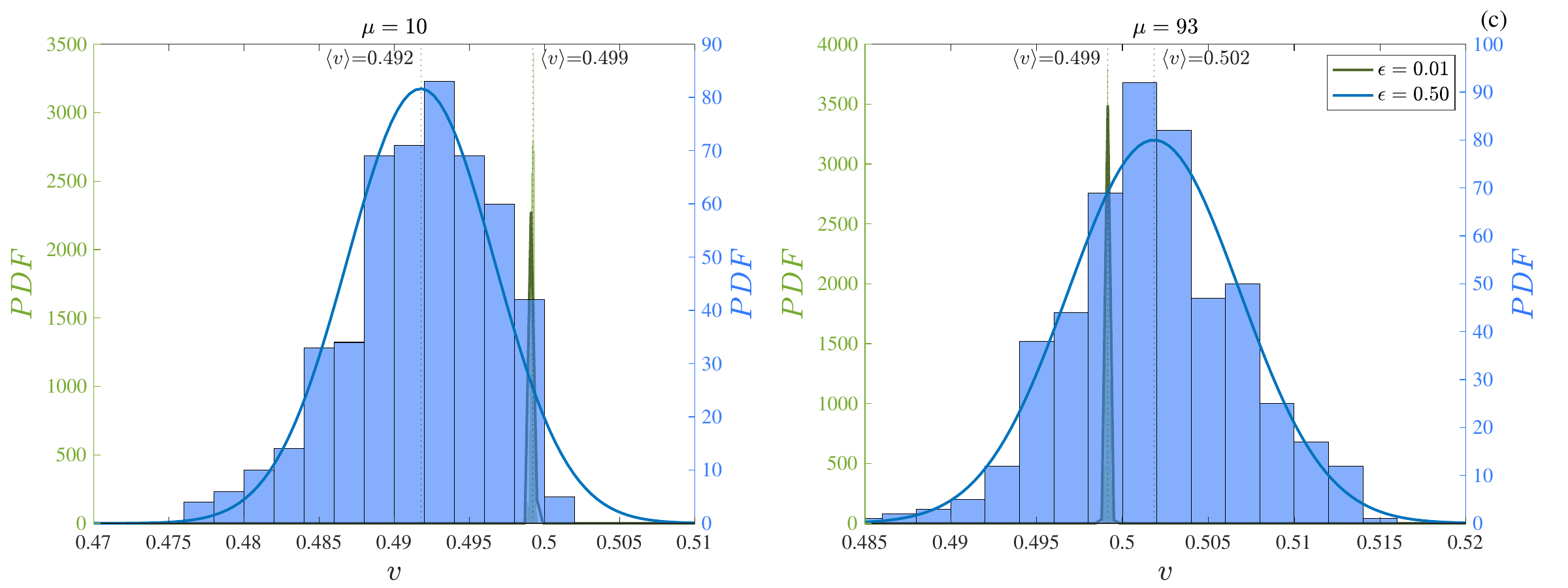}}
    \caption{Statistics of the chaotic speed enhancement under the Lorenz system. \textbf{(a)} Mean speed velocity $\langle v\rangle$ as a function of the disorder strength $\epsilon$ for different values of the bifurcation parameter: $\mu=28$ (strongly chaotic), $\mu=93$ (weakly chaotic), and $\mu=10$ (non evolving). \textbf{(b)} Phase space orbits and the largest Lyapunov exponent of the underlying Lorenz system. \textbf{(c)} Behaviour of the probability density functions (PDF) of the speed $v$ for $\mu=10$ (left panel) and $\mu=93$ (right panel), considered for different values of the chaos strength $\epsilon$.}
    \label{Fig:05}
\end{figure}

Figure~\ref{Fig:05} summarises our results, which confirms that the speed enhancement effect also appears in continuous-time chaotic systems. Figure~\ref{Fig:05}(a) evidences that the mean speed $\langle v\rangle$ increases with $\epsilon$ for $\mu=93$, which corresponds to a weakly chaotic evolving medium following the Lorenz system~\ref{Eq:Lorenz}. As a reference, we show in Fig.~\ref{Fig:05}(b) the largest Lyapunov exponent, $\mathcal{L}_{\max}$, and the corresponding phase space orbit of the underlying Lorenz system for different values of $\mu$, thus confirming that the time dependence of $\eta(x,t)$ is weakly chaotic for $\mu=93$, given that $\mathcal{L}_{\max}$ is small and positive.

As we observed in the case of the logistic map, highly developed chaos in the Lorenz system does not imply a stronger speed enhancement effect. Indeed, the Lorenz system is highly chaotic for $\mu=28$ [see Fig.~\ref{Fig:05}(b)]. Still, the speed-enhancement effect is weaker: the mean speed $\langle v \rangle$ rather than showing a sustained increase with $\epsilon$, slowly saturates towards a value around $\langle v\rangle\sim0.498$. Thus, evolving disorder at this level of chaos, rather than increasing the mean speed of waves, sustains its speed as the disorder's strength increases. We also show in Fig.~\ref{Fig:05}(a) and~\ref{Fig:05}(b) the case $\mu=10$, which corresponds to a non-evolving medium where the attractor of the Lorenz system is a stable fixed point. For this case, $\mathcal{L}_{\max}<0$, and the wave tends to localise with increasing values of $\epsilon$. The behaviour of the PDFs is shown in ~\ref{Fig:05}(c), which is consistent with our previous observations using the logistic map: the PDFs become wider for increasing values of $\epsilon$, whereas the mean speed of the wavefront increases (decreases) under chaotic (non evolving) disorder.

xº
The speed-enhancement effect reported in Fig.~\ref{Fig:05} was obtained for $t/\tau=10$, which gives the ratio of the proper time of the chaotic oscillations in the Lorenz system~\ref{Eq:Lorenz} to the time of the wave dynamics under Eq.~\ref{Eq:02}. This ratio suggests the notion of a time-scale competition, which states that to induce the speed-enhancement effect, the chaotic fluctuations must be fast enough to induce enough spatiotemporal complexity in the wave. The physical significance of this coupling of time scales is reminiscent of the emergence of Langevin dynamics arising from a chaotic (deterministic) kick-force in dynamical systems, as studied formally using iterative maps in Refs.~\cite{Beck1990, Beck1995, Beck1996, Williams2018}. Indeed, our deterministic dynamical system also exhibits a transition from small-scale non-Gaussian chaotic behaviour to a large-scale speed-enhancement behaviour when a suitable time scaling limit is applied \cite{Beck1990, Beck1995, Beck1996, Williams2018, Hilgers1999, Mackey2006}. It is important to remark that in our system, a similar time scaling limit also occurs in the case of chaos under the logistic map, which does not have a well-defined continuous counterpart. In Sec.~\ref{Sec:Logistic}, the coupling of the logistic map is performed at every time step of the numerical integration of Eq.~\ref{Eq:02}, which gives a rapidly fluctuating chaotic (logistic) perturbation to the wave. Thus, changing the time integration step $\Delta t$ inside the range of numerical stability and the CFL condition is equivalent to changing the time coupling between the wave and chaos. For relatively high values of the time step $\Delta t$, the logistic chaos fluctuates too slowly to induce enough complexity in the elastic wave to induce the speed enhancement effect. However, we expect to observe small quantitative changes to the speed enhancement effect decreasing $\Delta t$ below some critical value. This time-scale competition effect in our wave propagation model and its relation with the emergence of Langenvin dynamics from deterministic mappings will be studied elsewhere.

\section{Conclusions \label{Sec:Conclusions}}

In summary, we have shown that evolving disorder, random or chaotic, enhances the wave speed of elastic waves in one-dimensional systems under diagonal disorder. Using as a model a hyperbolic partial differential equation with space-time variations in the phase velocity, we have characterised the wave speed enhancement effect considering three types of statistics for the disorder: Gaussian (random), discrete-time chaotic (deterministic) and continuous-time chaotic (deterministic) distributions. Under frozen disorder, elastic waves localise as the disorder strength increases. However, we have shown that evolving random disorder tends to increase the speed of wavefronts as the disorder's strength increases. We have discussed similarities between the observed phenomenon and the hyper-transport of light observed by other researchers in random evolving media \cite{Levi2012}. Furthermore, we have discovered that chaotic (deterministic) evolving disorder induces the same speed-up effect. Thus, randomness is not a unique prerequisite to observing enhanced transport. We have characterised the chaotic speed-up effect considering complex signals generated from two paradigms of chaotic dynamics: the logistic map and the Lorenz system. We have shown that the maximum value of the ensemble-averaged speed reaches a maximum plateau within an interval of the bifurcation parameter capturing the transition to chaos. We provide evidence of the breakdown of chaotic speed-up at intervals of the bifurcation parameter corresponding to the periodic windows of the logistic map, thus supporting the hypothesis that the wave speed in such deterministic systems is purely enhanced by chaos and not by periodic signals composed of a large number of frequencies.

\section*{Acknowledgments} 
The work on ``evolving random disorder'' was started when L.T. and J.F.M. were associated scientists at Instituto Venezolano de Investigaciones Cient\'ificas (IVIC), Venezuela. The idea of ``chaotic disorder'' was pursued at Universidad Tecnol\'ogica Metropolitana and Universidad T\'ecnica Federico Santa Mar\'ia, Chile. 
M.A. thanks Agencia Nacional de Investigaci\'on y Desarrollo (ANID) for the financial support through
the FONDECYT postdoctorado grant No. 3240443.
L.T. thanks the ICTP,  INSA-Lyon, as well as the Inria's Beagle Team, LIRIS-CNRS for hospitality while part of this research was done.
L.T. was partially supported by the INSA Visiting Professor Fellowship.
J.F.M. thanks Dirección de Investigación at Universidad Tecnológica Metropolitana for the financial support. Project supported by the Competition for Research Regular Projects, year 2023, code LPR23-06, Universidad Tecnológica Metropolitana.
%

\appendix

\section*{Appendix}

\section{Derivation of the wave equation under diagonal disorder}
\label{App:1}
\setcounter{section}{1}

Elastic wave propagation in a one-dimensional medium can be described microscopically by a system of $n$ consecutive point-like particles of mass $m_j$ coupled by springs with a Hooke's constant $k_j$ ($j=1,\ldots,n-1$). Diagonal disorder corresponds to the situation where $k_j=k$ is constant and deterministic, whereas $m_j$ for $j=1,\ldots,n$ is given by a stochastic process. Let $\phi_j(\omega, t)$ be the displacement from equilibrium of particle $j$ at time $t$ in a realisation $\omega(t)$ of the random evolving medium. If such displacements are small compared to the inter-particle distance $a$, the system exhibits a linear response to external perturbations and the stochastic equation of motion for the $j$-th mass within the bulk of the material ($j=2,\dots,n-1$) reads
\begin{equation}
\label{Eq:App1}
a\rho_j(\omega, t)\frac{d^2\phi_j}{dt^2}=-\frac{2K_0}{a}\phi_{j}+\frac{K_0}{a}(\phi_{j+1}+\phi_{j-1}),
\end{equation}
where $\rho_j(\omega, t):=m_j(\omega(t))/a$ and $K_0:=ak$ are the microscopic mass density and stiffness of the medium, respectively. We can use some coarse-graining strategy to obtain a continuous (macroscopic) wave equation from Eq.~\ref{Eq:App1}. The use of integral representations through weighted space-time averages of particle quantities is one such alternative \cite{Murdoch2012}, which can be achieved by multiplying both sides of Eq.~\ref{Eq:App1} by a smoothing kernel of radius $h$, namely $f(x-x_j,h)$, and integrating over $x_j$ \cite{Murdoch1994, Glasser2001, Goldhirsch2002}. This type of coarse-graining formalism is known to solve problems related to discontinuities and the lack of derivatives in many stochastic processes, such as the case of fractional Brownian motions (as shown in Mandelbrot's seminal work on Ref.~\cite{Mandelbrot1968}). Here, we assume that the kernel $f$ has a compact support domain $\mathcal{S}(x)$ of radius $h$ around $x_j$, i.e. the function $f$ vanishes if $|x-x_j|\geq h$ \cite{Marin2014, Petit2014}. Thus, the left-hand side of Eq.~\ref{Eq:App1} becomes
\begin{equation}
\label{Eq:App2}
a\int_{\mathcal{S}(x)}\rho_j(\omega,t)\frac{d^2\phi_j}{dt^2}f(x-x_j,h)dx_j=a\langle\rho(x,t)\partial_{tt}\phi(x,t)\rangle_{h},
\end{equation}
where $\langle\cdot\rangle_{h}$ denotes the continuum function obtained after coarse-graining with a smoothing function of radius $h$. Similarly, the right-hand side of Eq.~\ref{Eq:App1} after coarse-graining and using a Taylor expansion for $\phi_{j\pm1}-\phi_j\gg a\forall j$, becomes
\begin{equation}
\label{Eq:App3}
\frac{K_0}{a}\int_{\mathcal{S}(x)}(\phi_{j+1}-2\phi_{j}-\phi_{j-1})f(x-x_j,h)dx_j=aK_0\langle\partial_{xx}\phi(x,t)\rangle_{h}.
\end{equation}
Equating Eqs.~\ref{Eq:App2} and~\ref{Eq:App3}, setting $c(x,t):=\sqrt{K_0/\rho(x,t)}$, we obtain a hyperbolic partial differential equation of the same form as Eq.~\ref{Eq:02}.

\section{Calculation of the Lyapunov exponent}
\label{App:2}

The Lyapunov exponents in Fig.~\ref{Fig:03}(b) were estimated using the \emph{derivative method}, as described in Ref.~\cite{Lynch2004}. This method is particularly effective for simple iterative maps like the logistic map. Specifically, we computed the Lyapunov exponent $\mathcal{L}(r)$ at each value of the bifurcation parameter $r$ using the formula:
$$
\mathcal{L}(r) = \frac{1}{N} \sum_{n=1}^{N} \ln \left| \left. \frac{\mathrm{d} f_r}{\mathrm{d} \eta} \right|_{\eta = \eta_n} \right|,
$$
where $N = 100999$ and $f_r(\eta) = r\eta(1 - \eta)$, as defined by the logistic map. The exponents were computed near the attractor, thus ensuring that they are independent of the initial conditions of the orbit.

Additionally, alternative methods such as the Jacobian matrix method, Wolf's algorithm, and its modifications (see Ref.~\cite{Steeb2014}) were considered. For the logistic map, all these methods yield results consistent with those shown in the inset of Fig.~\ref{Fig:03}(b). For the continuous-time dynamical system considered in this work, namely the Lorenz model, we employed the celebrated Rossesntein's algorithm to compute the largest Lyapunov exponent of Fig.~\ref{Fig:05}(b), as described in Ref.~\cite{Rosenstein1993}.

\section*{References}
\bibliography{iopart-num}


\end{document}